\documentclass[12pt]{article}
\usepackage{amsmath}
\usepackage{amssymb}
\usepackage{graphicx}

\usepackage[a4paper,left=28mm,right=25mm,top=30mm,bottom=30mm]{geometry}

\RequirePackage[OT1]{fontenc}
\usepackage{dcolumn,warpcol}  
\RequirePackage{amsthm,amsmath,amssymb}
\usepackage{graphicx}\graphicspath{{pics/}}

\usepackage{latexsym}
\usepackage{arydshln,booktabs,bm}
\usepackage{array}

\usepackage{here}
\usepackage{bm} % bold math

\usepackage{lscape} % page rotation

\usepackage{natbib}
\usepackage{array}
\renewcommand{\baselinestretch}{1.2}

\usepackage{fancyhdr}
\usepackage{bm} % bold math
\usepackage[latin1]{inputenc}
\usepackage{color}
\usepackage{graphicx}
\RequirePackage[OT1]{fontenc}
\usepackage{latexsym}
\usepackage{booktabs,rccol,flafter,arydshln,warpcol}
\usepackage{natbib}
\RequirePackage{amsthm,amsmath, natbib}

\usepackage{here}

\newcommand{\ci}{\mbox{\protect{ $ \perp \hspace{-2.3ex}
\perp$ }}}

\newcommand{\n}[0]{\hspace*{.35em}}

\newcommand{\nn}[0]{\hspace*{.7em}}

\newcommand{\ful}{\mbox{$\, \frac{ \nn \nn \;}{ \nn \nn
}$}}

\newcommand{\fla}{\mbox{$\hspace{.05em} \prec
\!\!\!\!\!\frac{\nn \nn}{\nn}$}}

\newcommand{\fra}{\mbox{$\hspace{.05em} \frac{\nn
\nn}{\nn
}\!\!\!\!\! \succ \! \hspace{.25ex}$}}

\newcommand{\dal}{\mbox{$  \frac{\n}{\n}
\frac{\; \,}{\;}  \frac{\n}{\n}$}}

\makeatletter
\renewcommand\section{\@startsection{section}{1}{\z@}%
{-3.25ex\@plus -1ex \@minus -.2ex}{1.5ex \@plus .2ex}%
{\normalfont\large\bfseries}}
\renewcommand\subsection{\@startsection{subsection}{2}{\z@}%
{-3.25ex\@plus -1ex \@minus -.2ex}%
{1.5ex \@plus .2ex}%
{\normalfont\normalsize\bfseries}}
\renewcommand\subsubsection{\@startsection{subsubsection}{3}{\z@}%
{-3.25ex\@plus -1ex \@minus -.2ex}%
{1.5ex \@plus .2ex}%
{\normalfont\normalsize\bfseries}}
\renewcommand\paragraph{\@startsection{paragraph}{4}{\parindent}%
{3.25ex \@plus1ex \@minus .2ex}%
{-1em}%
{\normalfont\normalsize\bfseries}}
\makeatother
\renewcommand{\baselinestretch}{1.2}

\newcommand{\odds}{\operatorname{\cal O}}

\begin{document}

\begin{center} {\large \bf  Case-control studies  for rare diseases: improved estimation of several risks 
and  of feature dependences} \\ \end{center}

\begin{center} {Nanny Wermuth, Chalmers Technical University, Gothenburg and IARC Lyon;
Giovanni Marchetti, Department of Statistics, Florence; Graham Byrnes, IARC Lyon}\\
\end{center}

\noindent{ \bf Abstract.} {\em To capture the dependences of a disease on  several
risk factors, a challenge is to combine model-based estimation with evidence-based arguments.
 Standard case-control methods allow   estimation of  the dependences of a rare disease on several regressors via  logistic regressions.  For  case-control studies, the  sampling
design leads to samples from two different populations and
 for the set of regressors in every logistic regression,  these  samples are then mixed and taken as given observations.   But,  it is the differences in
 independence structures of regressors for cases and for controls that can improve  logistic regression estimates  and  guide us to the important feature dependences that are specific to the diseased. A   case-control study on laryngeal cancer is used as  illustration.}
\section{Introduction}
\subsection{Motivation and goals}

Logistic regression models  the dependence of a binary response variable on a set of regressor variables. These models have proven to be extremely successful tools in epidemiological and medical research, where the response is for instance   survival or  a  disease diagnosis.  In these models, the parameters measuring dependence are the log-odds ratios.
And,  in  any two-way classifications of a response and a regressor, only  functions of the odds-ratios
remain unaffected   by changes in the margins and hence   by 
 changes in the sampling schemes that allow to fix the overall count of the studied individuals, the row sums or the column sums; see Edwards  (1963).

The  variants of these models with  exclusively categorical regressors  are named 
 logit regression; see Fienberg (1980, 2007).   For logit regressions,   goodness-of-fit tests compare  how well  the estimated counts  agree with the  observed counts.  For  general logistic regression instead, in which some of the regressors are quantitative and without replication for all  level combinations, there is no similar formal test of goodness-of-fit. In such logistic regressions,  a poor quality of the fitted values shows in increased standard deviations of
the estimates but these give no direct  pointers to locations of  poor fit. Also, for some research questions,  originally quantitative measures may not be of direct interest so that logit regressions can be focused on.

One may see it as an advantage  of a model if it estimates in regions of sparse data by borrowing strength 
from portions of the data with many observations, however there is  the danger  that
conclusions are extrapolated to situations for which no data support is available, or to put it differently,
for which evidence-based arguments  are lacking since  parameter estimates and  statistical tests cannot be complemented by convincing basic data descriptions.

A  special situation arises  for case-control data. By design,  such data are always samples from two different populations. There is one population  of the cases, the individuals  who have been diagnosed to have  the disease under study and  another population of  the controls, the individuals   without the disease.  

For any rare disease, the cases constitute a tiny fraction of the general population in a given region at the time of study, so that for general public health decisions, the distribution of risk factors 
among the controls is of main interest. But for contemplating  for instance a subsequent prospective study or the implementation of a small  preventive  screening program for those at highest risk,  the main focus of  interest   are those combinations of risk factors and of other features that  summarize best how the cases differ from the controls. 

 For case-control data, one  conditions with each regression model on a mixture of data from two populations. The  mixing proportions of  one  case to several controls are typically determined by cost  considerations  and,  for  rare diseases, they  never define a sample of a real population. Dependences among regressors that exist either in the population
of cases or in the population of controls may appear enlarged,  diminished or get cancelled  with the mixed sample values that are taken in regressions as the given observations of  the regressors. 

Therefore, we supplement the traditionally used  logistic regressions for case-control data in two main ways. First, we obtain subgroups of cases and controls  that are relevant for the main research questions and
that are  comparable in related, important  features.  Second, we carry out  separate analyses  for cases and for controls  and exploit differences in dependence structures to obtain improved, smoothed estimates of odds-ratios and to identify  important feature combinations that distinguish  cases  from controls.\\[-8mm]

\subsection{Background information to laryngeal cancer and to regression graphs} 
 Laryngeal cancer is currently  the second most common cancer of the
respiratory tract. Several risk factors have been reported in the literature. The  risk in males increases with the amount and duration of tobacco consumption;
see e.g. Zatonski et al.  (1991), with regular  alcohol consumption and with decreasing age at start of smoking,
see e.g. Talamini et al.  (2002).  It  decreases with an increasing time since ceasing to smoke; see IARC Monograph 83 (2003)
and  with only light inhalation when smoking cigarettes; see Ramroth et al. (2011).  Differential susceptibility
for  laryngeal cancer in men and women may be related to hormones being involved in the carcinogenic process and to
a gender-specific predisposition to develop laryngeal cancer; see Chen et al. (2011).

However, the understanding of the carcinogenic process and of the joint effects of alcohol and tobacco
consumption is  still limited.    We focus mainly on the combination of  these
  two well-established risk factors  for laryngeal cancer when there are  high exposures to both and when some of the
  intrinsic variables are  comparable; see Berrington and Cox (2007) for a discussion of effects of  intrinsic variables, which are characteristic of the study participants that  cannot be modified by intervention, and
  see Cornfield et al. (1959), Cox and Wermuth (2003) for statistical aspects of causal relations.

  Conditional independences identified separately in the sample of cases and  in the sample of controls, lead to   well-fitting, simplified dependence structures, to improved estimates of  risks and to  profiles of persons being at highest risk. 
  
  Regression graphs capture both independences and sets
of   directly important explanatory variables;
see Cox and Wermuth (1993), Wermuth and Sadeghi (2012).  Here, laryngeal  cancer is the primary  response of interest.
Tobacco and  alcohol consumption represent the two components of a  joint  response  of secondary interest, which  may depend  on additional  features of the 
participants that are considered to be also relevant for the occurrence of this type of cancer. The two components of the joint response are in turn
joint risk factors for  laryngeal cancer or, to put it differently, they are explanatory for the response of main 
interest.

Thus for a general population in which laryngeal cancer is rare, a first ordering of the variables  in Figure 1  starts with the variable
of main interest on the left, shows alcohol and tobacco consumption as intermediate between laryngeal  cancer 
and  several background variables, listed  on the right, most of them being intrinsic features. 

\begin{figure}[H]
\centering
\includegraphics[scale=.52]{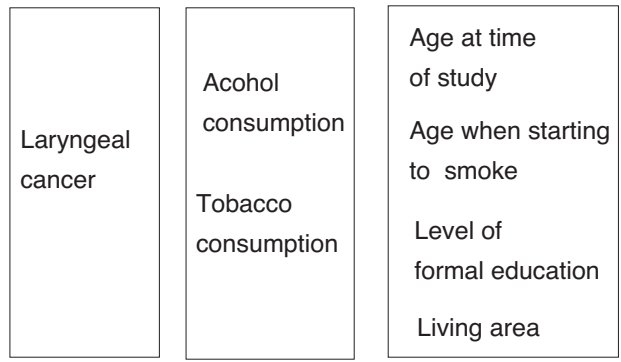}
\caption{\small A first ordering of  several variables relevant for laryngeal cancer.}
\label{figlabin}
\end{figure}
%\n \\[-20mm]
A regression graph has nodes representing variables and an edge coupling a node pair for dependences that are important in generating the joint distribution. An arrow, $i \fla j$, points from  a parent  node $j$ to its offspring node  $i$, a dashed line, $i\dal j$, connects any two  response nodes $i,j$ if they are dependent given their   joint set of parents and a full line, $i\ful j$ couples two background nodes $i,j $ if they are dependent given all 
other background nodes.

 For a regression graph constructed by statistical analysis from a case-control study, the arrows pointing to the disease result from estimated 
log odds-ratios and, for a general population of a rare disease, the dependence structure of all its potential regressors from an analysis of  the controls alone. Additional insights can be gained by using more explicitly  that, by the  case-control sampling design,  one has  conditioned  on each level of a binary response.  

One consequence for all important regressors of the main response is  that  their dependence structures
show, separately for the cases and for the controls, in  graphs of only full lines, which have been named their induced concentration graphs; see Wermuth (2011). Different cliques  for the same node subsets in  these two types of graphs point to those dependence substructures of the  features under study where the cases  differ mainly from the controls,  the disease diagnosis set again aside.

An induced concentration graph is 
Markov equivalent to a  regression graph in the same node set  and in the same set of edges  if and only if it can be oriented such that no 
collision {\sf V} results, that is no three-node, two-edge subgraph of either one of the types
$$  i \fra o\fla j, \nn \nn   i\dal o \fla j, \nn \nn i\dal o\dal j\,;$$
see Wermuth and Sadeghi (2012). Thereby, the first ordering of the variables is taken into account, the order that captures the available  knowledge about how the  joint distribution is generated.

Some difficulty  of variable ordering arises in  case-control studies when age at  the time of study  does not have the same meaning for cases and controls.
For
the general population, studied with the sample of the controls, age  is clearly the  intrinsic variable of an individual. 
For the cases however,  age is  also a proxy for the length of exposure to risks and hence for a higher morbidity status and as such a  consequence of  exposure instead of a background variable.
\\[-8mm]

\subsection{Background information on the case-control study used as illustration}

We use data from a Polish study of 249 males with exclusively glottic or subglottic tumors diagnosed in 1986 and 1987 in Lower Silesia;
see Zatonski et al. (1991).   The incidence rate of laryngeal cancer is reported for Lower Silesia as 35 in $100\,000$ for males aged 34 to 65  in the years 1984 to 1987; see Parkin et al. (1997).

 Of the laryngeal  cancer cases in this study, 80\% were blue collar workers, 79\%
lived in an urban environment, that is in towns of more than $10 \, 000$ inhabitants, and 76\% had less than 8 years of formal schooling.
 In this context,  gender,  age when  cigarette smoking was started, living area are directly important intrinsic features while  the level of formal schooling is also a  crude indicator for the standard of living of blue
 collar workers in Poland  at the time.  
 
 For this study, controls were selected from the electoral roll, with stratification
 by five-year age group and living area, rural versus urban. This stratification was used to ensure  that the sampled proportions
 conform to those in the population, since the authors state that `the age distribution
among controls is closely representative of that of the
general population within the age range 25 - 65 years,
whereas the cases tend to be older.'   
Slightly fewer than 5 controls were recruited per case. Response rates
 in cases and controls were high at 88\% and 94\% respectively.
This resulted in the participation of 249 cases and 965 controls, a total of 1214 men in the 
original study group.

\subsection{Obtaining comparable relevant features of cases and controls}

For a reliable estimation of the dependence of a 
disease on two or more specific risks, the choice of controls is critical. The studied groups of cases and  controls 
need to be comparable with respect to other  features that may distort these dependences when they are ignored. Such
features have been  called possible direct confounders; see Wermuth and Cox (2008).

In the literature on laryngeal cancer,  tobacco consumption is typically described as a more important risk than alcohol consumption. We expect this to be different for high exposures and
with the given data of the  Polish study, we  contrast high versus extreme exposures. The latter are    long-term heavy vodka drinking and long-term heavy  cigarette smoking.

The features considered  as background variables,  listed  in the right-hand box of Figure 1, are  possible direct confounders.
In rural more than in urban Polish communities,  heavy vodka drinking and heavy cigarette smoking may have been socially accepted.  Cumulated exposure to these risks may  increase with age at time of study  and with the age when cigarette smoking was started.
Poor nutrition leads to poor health and higher morbidity and may be more likely  the lower the level of formal education.

To  obtain comparable groups with respect to these possible confounders,  we   use the same constraints on the  cases and controls of the  original study group.
 Every  selected man smoked on average 10 or more cigarettes per day, without any long non-smoking periods and without
 ceasing to smoke more than 4 years prior to the time of study. He
 started smoking 10 or more years before the time of study, at an age of 26 years or younger.   Further, every selected man was at least
 33 years old and had less than 12 years of formal schooling since, otherwise, there would have been not one comparable case in the rural communities.

The  constraints taken together exclude more cases than controls, leaving 31\% cases (48  of 156) in rural areas and
37\% cases (156 of 424) in urban areas. Combined, there are 35\% cases (204 of 580) in the selected study group  compared
to  only 7\% cases (45 of 634) in the subgroup excluded from the original study group.

We define seven binary variables  from the  given, available  raw data.  
With $n=580$
study participants and  seven binary variables, the classification into $ 2^7=128$
level combinations gives  27\% empty cells. Hence any estimation
can  be successfully combined with  evidence-based arguments only if the data support  many simplifying  
independences or  if a dependence model has been shown to be appropriate in a previous, larger  study. 

The binary variables are  $L$, laryngeal cancer (1:=case, 0:=control),  $V$, heavy vodka drinking (1:=yes, regularly since 5 or more years, 0:=no),
 $C$, heavy cigarette smoking (1:=yes,  more than 20 cigarettes per day, 0:=no, 10 to 20 cigarettes per day),
 $A$, older age group at time of study (1:=yes,  51 to 65 years, 0:=no, 33 to 50 years),
 $E$, lower level of formal education (1:=yes, less than 8 years, 0:=no, 8 to 11 years),
 $S$, starting age for cigarette smoking (1:=18 years or older but at least  ten years before  the time of study , 0:= 17 years or younger),
 $R$, region of living, (0:=rural,  1:=urban).
 Except for variables $S$ and  $R$,  level 1 of the other four  explanatory variables 
 represents what is expected to be a higher level of risk for laryngeal cancer  than level 0.

  Tables 1 and 2 give the typical features for  the  four subgroups defined by the region of living, $R$, and the type of cigarette smoking, $C$. 
 
\begin{table}[H]
\caption{Features of the selected regular and heavy smokers  in  rural Polish communities}\n \\[-3mm]
\centering \small
\begin{tabular}{l rrrr l rrrr }
\hline \\
[-3mm] &  \multicolumn{4}{c}{regular smokers, $n=103$}&\n&\multicolumn{4}{c}{heavy smokers, $n=53$}\\
feature & min&max& mean& std& \n & min& max& mean& std \\
\hline \\
[-3mm] case, 0:=no, 1:=yes& 0 &1 &  22\%& \n& & 0& 1& 47\%\\
yrs of drinking hard liquor & 0& 30& 2.3& 6.62&&0 & 30 & 7.4&10.74\\
 $> 4$ yrs heavy drinking, no/yes& 0 & 1& 12\% &&& 0& 1& 36\%&\\
 $< 8$ yrs education no/yes&0&1&68\%&&&0&1& 74\%&\\

 age in yrs at time of study&33&64&48.8&10.22&&33&64&49.6&8.83\\
 age $>50$ yrs, no/yes& 0 &1 & 49\%&&&0&1& 49\%&\\
 age in yrs  smoking started&10&26&18.1&2.58&& 12&26&18.2& 3.09\\
 start smoking $>17 $ yrs, no/yes& 0& 1& 67\%&&&0&1&66\%\\
 av. \# cigarettes per day & 10&20&17.2&3.16&& 21 &60& 30.0& 8.29\\
  [1mm]  \hline \vspace{-3mm}
 \end{tabular}
 \end{table}
 
 By Tables 1 and 2,  the four
groups are comparable with respect to the intrinsic feature 
age  when cigarette smoking was started.   Within living areas,  there are roughly similar percentages of  participants  with a low level of formal education. The average number of cigarettes smoked per day is comparable
for rural and urban regular smokers (17.2  and 17.7)  as well as for rural and urban heavy smokers
(30.0 and 30.1).

  With the four group sizes being quite different, the  much higher average level of tobacco consumption for heavy
 smokers than for regular smokers requires conditioning on smoking, while  the  longer exposure to heavy vodka drinking
 in urban compared to rural regular smokers requires conditioning on region when studying  the effects
 of heavy vodka drinking in addition to heavy smoking.

The chances of seeing cases of laryngeal cancer range from 22\% to 47\% in the four groups. After merging the
two groups of smoking given the regions or by merging the two regional groups given the levels of smoking,
the proportion of cases would only vary between 31\% and 37\% for the  two regions  and between  31\% and 47\%
for the two groups of cigarette smokers. 
\begin{table}[H]
\caption{Features of the selected regular and heavy smokers  in  urban Polish communities}\n \\[-3mm]
\centering \small
\begin{tabular}{l rrrr l rrrr }
\hline \\[-3mm]
&  \multicolumn{4}{c}{regular smokers, $n=310$}&\n&
  \multicolumn{4}{c}{heavy smokers, $n=114$}\\
feature & min&max& mean& std&\n&
 min& max& mean& std \\
\hline \\[-3mm]
 case, 0:=no, 1:=yes& 0 &1 &  33\%& \n& & 0& 1& 46\%\\
  yrs of drinking hard liquor & 0& 46& 4.4& 10.22&&0 & 46 & 6.6&11.77\\
 $> 4$ yrs heavy drinking, no/yes& 0 & 1& 17\% &&& 0& 1& 28\%&\\
$< 8 $ yrs education no/yes&0&1&57\%&&&0&1& 61\%&\\
 age in yrs at time of study&33&65&49.2&9.66&&33&65&49.0&8.79\\
 age $>50$ yrs, no/yes& 0 &1 & 53\%&&&0&1& 46\%&\\
 age in yrs  smoking started&7&26&18.5&2.71&& 9&26&18.0& 2.80\\
 start smoking $ > 17 $ yrs, no/yes& 0& 1&69\%&&&0&1&65\%\\
 av. \# cigarettes per day & 10&20&17.7&2.99&& 21 &60& 30.1& 7.61\\
  [1mm]  \hline
 \end{tabular}
 \end{table}
Thus by merging, one would give up the opportunity of
investigating strong contrast in the numbers of cases to controls and  
 marginal estimates could be distorted  estimates of the
conditional effects of heavy vodka drinking on laryngeal cancer.  \\[-8mm]

 \section{Measures of  marginal, conditional  and overall dependence}
 
\subsection{Relevant recent results for different measures of dependence}

For data obtained with a case-control design,  the  relation of counts for cases and controls
depends on the number of selected controls per  case and on the selection criteria.
As one consequence, any observed percentage of cases  does not translate directly
  into an estimate of the probability that the illness occurs.

When
 $n_{\rm 1y}$  denotes the count of cases   and $n_{\rm 0y}$ the
 count of controls  at  high exposure to  a risk factor (y:=yes),   then the  odds for cases versus controls
 are $n_{\rm 1y}/n_{\rm 0y}$, while  otherwise they are  $n_{\rm 1n}/n_{\rm 0n}$ for  n:=no.
 For the study group of $n=580$ men,  the following  table of  counts and percentages gives as  observed odds-ratio  (odr) for
 laryngeal cancer, $L$,  and  heavy  vodka consumption, $V$, as
 $${\rm odr}(LV)= (n_{\rm 1y}n_{\rm 0n})/(n_{\rm 0y}n_{\rm 1n})=(349 \times  88)/(116 \times 27)=9.8.$$
 With a standard deviation of $\sqrt{1/349 + 1/88 + 1/116 + 1/27} = 0.245$  for $\log {\rm odr}(LV)$, 9.8 represents a highly significant deviation from 1, the expected value  of the odds-ratio when two binary
 variables  are independent. 
 
  The observed odds-ratio for an illness and a risk factor,  for instance ${\rm odr}(LV)$, takes  on a value of  1 if and only if the
  percentages of cases  agree in the two categories of exposure. In addition, an odds-ratio approximates  a   relative
  risk for rare events. In this latter case,  the odds  expressed in terms of probabilities, say  as $\pi/(1-\pi)$,
  are roughly the same as  the probability for the rare event, $\pi$, itself. 
  Thus, if a disease is rare in the population,  odds ratios  estimated in  case-control studies,  estimate relative risks in a population; see also Prentice and Pike (1979).

  %On the other hand,  provided these are sampled independently of the exposures of interest. 

In Table 3, the observed odds-ratio
 ${\rm odr}(LV)=9.8 $ differs substantially from the ratio of the observed percentages, 77/25= 3.1 and from the observed risk difference, $77-25=55$, the  change in percentage points. However, the three measures always agree regarding a positive dependence of $A$ on $B$ given $C,Z$ when $A,B,C$ are binary variables and $Z$ is  vector background variable which may be a combination of categorical variables, of quantitative variables or consist of both types. 
 
  \begin{table}[H]
 \caption{Counts  of laryngeal cancer  cases for given levels of vodka consumption}
 \begin{center}
{ \begin{tabular}{l  rrr rr}
\hline
&\multicolumn{3}{c}{$V$, heavy vodka consumption }\\
\cline{2-4}
$L$, laryngeal cancer &no  & yes&&& sum\\
\hline
0:=controls & 349    &   27& &&376\\
1:=cases &  (25\%) 116& (77\%) 88&&& (35\%) 204\\
\hline
sum & 465 & 115&&& 580\\
\hline
odr$(LV)$& \multicolumn{2}{c}{\nn \n \nn 9.8}&&&\\
 \hline \end{tabular}}
\end{center}
\end{table}

To be more precise, we denote  a probability $\Pr(A=i, B=j, C=k, Z=z)$ by $\pi^{ABCZ}_{ijkz}$ and marginal probabilities  by summing over some of the levels  $+$, so that for instance $\pi^{+BCZ}_{jkz}=\sum_i \pi^{ABCZ}_{ijkz}$ and the conditional probability of $A=1$ given $B,C,X$ is defined by $\pi^{A|BCX}_{1|jkz}=\pi^{ABCZ}_{1jkz}/\pi^{+BCZ}_{jkz}$. 

If we denote further the odds for $A=1$ given $B=1$ and arbitrary levels of $C,Z$ by
$\odds\!{\rm d}(A|BCZ)=\pi^{ABCZ}_{11kz}/\pi^{ABCZ}_{01kz}$, and for  $A=1$ given $B=0$ by $\odds\!{\rm d}(A|\bar{B}CZ)$, then the  odds-ratios of $AB$ given $C,Z$  are  $\odds\!{\rm dr}(AB|CZ)=
\odds\!{\rm d}(A|BCZ)/\odds\!{\rm d}(A|\bar{B}CZ)$, the relative risks  for $A=1$ comparing $B=1$ to $B=0$ given $C,Z$ are $\pi{\rm rr}(A|BCZ)=\pi^{A|BCZ}_{1|1kz}/\pi^{A|BCZ}_{1|0kz}$
and  the following three statements of positive conditional dependences of $A$ and $B$ can be shown to equivalent; see also Theorem 1 of Xie, Ma and Geng (2008):
\begin{eqnarray} \label{equiv}
  \pi{\rm rr}(A|BCZ)&>&1\\
   \odds\!{\rm dr}(AB|CZ)&>&1  \\
 \pi^{A|BCZ}_{1|1jz}-\pi^{A|BCZ}_{1|0jz}&>&0\,.
 \end{eqnarray}
The same types of relation hold for observed instead of expected contingency tables,
as has been illustrated with Table 3. Equality in the three equations corresponds to  conditional independence of $A,B$ given $C,Z$,  written compactly as $A\ci B|CZ$.  
 %In  the next section, we  alert to 
% the different types of conditions to see in a marginal table the same dependence as in equal conditional  odds-ratios compared to equal conditional relative risks.

If  the conditional independence $A\ci X|B$ holds, then the conditional density given $B$ factorizes accordingly, as $f_{AX|B}=f_{A|B}f_{X|B}$ and any measure of dependence of $A$ on $B$ given $X$ is unchanged after  marginalising over $X$.

Given  this result, one knows that two equal conditional odds-ratios  for $AB$ given a binary  variable $C$ coincide with the marginal odds-ratio,  $\odds\!{\rm dr}(AB)$, if $A\ci C|B$.  But, there is an additional sufficient condition, since 
the conditional odds, $\odds\!{\rm d}(A|BC)$,  relate   via conditional probabilities to the 
 marginal odds as
 $$ \odds\!{\rm d}(A|BC)=\odds\!{\rm d}(A|B)(\pi^{C|AB}_{1|11}/\pi^{C|AB}_{1|01})\, $$
 Therefore, 
 $  \odds\!{\rm dr}(AB|C)=\odds\!{\rm dr}(AB)$ if the second term on the right-hand side does not depend on $B$,   that is if $C\ci B|A$. These two results together  provide 
 a direct proof of simple collapsibility of equal conditional odds-ratios for binary variables $A,B$  over a binary variable $C$, that is 
 \begin{equation}  \{\odds\!{\rm dr}(AB|C=0)= \odds\!{\rm dr}(AB|C=1)= \odds\!{\rm dr}(AB) \}\iff  
  (A\ci C|B \text{ or } B\ci C|A)\, ;\end{equation} 
  see also Whittemore (1978).

 The 
same type of conditions as in (4)  are sufficient but not necessary for simple collapsibility of  equal odds-ratios when  conditioning is  on more variables or on categorical variable with more than two levels.  
 In the framework of graphical models,  these are discussed  by  Didelez, Kreiner and Keiding (2010) and  Barenboim and Pearl (2011)   as conditions for controlling selection bias for  odds-ratios when there is   outcome dependent sampling. However,   conditions for  simple collapsibility of equal relative risks given $C$   are  quite different from those for equal odds-ratios given $C$; see also Geng (1992). 
 
 If $B\ci C$, then the marginal relative risk of $A,B$ is an average of the conditional risks of $A,B$ given both levels of $C$ with positive weights adding to one:
 $$  \pi{\rm rr}(A|B)=\{ \alpha \, \pi{\rm rr}(A|BC) + \beta\, \pi{\rm rr}(A|B\bar{C})\}/(\alpha+\beta)\,, $$
 where $\alpha=\pi^{++C}_{1}\pi_{1|01}^{A|BC}$ and $\beta=\pi^{++C}_{0}\pi_{1|00}^{A|BC}$.
 Thus, positive conditional dependences always lead to a positive dependence after marginalizing over $C$ and if the conditional relative risks given $C$ are equal, then  they also coincide with the marginal relative risk for $A$ given $B$ that is
 \begin{equation}
 \{ \pi{\rm rr}(A|BC)= \pi{\rm rr}(A|B\bar{C})= \pi{\rm rr}(A|B)\} \iff   (A\ci C|B \text{ or } B\ci C)\, .\end{equation}

There can be strong differences in conditional odds-ratios in spite of equal conditional relative risks
and there can be strong differences in conditional relative risks in spite of equal conditional odds-ratios.
Both results are useful  for conclusions to be drawn  from case-control studies, as shown here in Sections 3.5 and 4.
 
Possible effect reversal also needs  attention. It is well-known that strongly dependent  explanatory variables $B,C$  can lead to a marginal positive dependence of $A$ on $B$ being reversed after conditioning on $C$, or vice versa, that conditionally positive dependences of $A$ on $B$ given $C$ can be reversed when $C$ is ignored. This is relevant only  when variables $B,C$ are two strongly dependent explanatory variables and there is, in addition, a qualitatively similar dependence of  response $A $ on $B$  at  the given levels of $C$. This can be deduced from the equivalent ways of expressing positive dependence in equations (1) to (3) in combination with  Theorem 1  of  Vellaisamy (2012).

 In case-control studies, such  situations may arise   in  logistic regressions with the disease as response, for instance, when two important regressors  measure essentially the same risk,  or when a   strong dependence  is  induced between two important risk factors by the specific  way in which  samples of cases and controls are mixed in the given data.\\[-8mm] 
 
 \subsection{The overall effects of the six observed, possibly explanatory variables}

With  Table  4, we report the overall odds-ratios for response $L$.
It shows   the variable pair in column 1, the observed overall odds-ratios for each variable pair  in column 2 and the corresponding likelihood-ratio statistics for independence in column 3.  Pearson's chi-square statistics
  for pairwise independence are displayed in column 4,  Pearson's correlation coefficients for the binary variables   in column 5 and 
 Pearson's correlation coefficients  as computed  for the original binary  and quantitative variables in column  6.\\[-8mm]

\begin{table}[H]
 \caption{Observed pairwise marginal dependences for $n=580$}
 \begin{center}
\begin{tabular}{l  rrrrr}
\hline
variable& odds-& likelihood & Pearson's&\multicolumn{2}{c}{ Pearson's correlation}\\
pair& ratio & $\chi^2$&   $\chi^2$& binary& raw data\\
\hline
$ (L, V)$&        9.8 &       104.5 &        107.6&          0.43   &       0.43\\
$ (L, C)$&       2.0    &     13.4 &         13.7&         0.15  &        0.18\\
$ (L, A)$&          3.8 &         54.5 &         53.2&        0.30   &       0.31\\
$ (L, E)$&         3.7 &         46.2&         43.9&         0.28     &    0.28\\
$ (L, S)$&          1.4 &          2.5 &          2.5&          0.07         &$-$0.01\\
$ (L, R)$&          1.3 &          1.8 &       1.8 &          0.06      &    0.06\\[1mm]
          \hline
          \end{tabular}\\[2mm]
          \small Note that variable $E$ has been coded to  have 1:= lower level of education
\end{center}
\vspace{-5mm}
\end{table}

The last two columns  of the table show how little the correlation coefficients of  the defined binary variables differ in size 
  from correlations with the quantitative measurements in the raw data.
In addition,
the two types of chi-square statistics agree extremely
  well for the given set of data.
  In general,  this is by no means the case since the likelihood ratio statistic is
  based on the odds-ratio and Pearson's chi-square statistic is instead a monotone function of the correlation
  coefficient of two binary variables.

  A high agreement of the likelihood chi-square and  of Pearson's chi-square can  be expected
  when the binary data  have  two special properties:
  (1) none  of the directly observed  binary variables concerns a rare event
  and (2) the cutoff-points for the defined binary variables do not generate rare events.
  The latter  happens, in particular, when the cutoff-point is near the median. This holds here since the  observed 
  percentages at level one are  for $L,V,C,A,E,S, R$, respectively, in percent
 $  35.2,   19.8,   28.8,   50.5,   61.2,   67.4,   73.1.$

  The relevant, more general  result  is due to  Cox (1966) who shows
 that  the estimated slope in a simple logistic regression does not differ much from a linear regression coefficient
 whenever the   observations at each of  extreme  levels  of the response  do not correspond to a rare event, that is  to percentages smaller than $10\%$, say.

 %And, the coefficient of dependence in  a simple logistic regression with a binary regressor, that is in a simple logit regression,
% is  the  logarithm of the odds-ratio.

   With $r$ denoting  a correlation coefficient of two binary variables  and $n$ the sample
  size,  the value of Pearson's chi-square  statistic  is  $nr^2$; for a proof see for instance Wermuth and Streit (2007), p. 344.  
  Expressed differently, for $n=580$, an  absolute value of these observed correlation  coefficients  larger than
 $ 0.11=\sqrt{6.63/580}$ is statistically significant at  a $0.01$ level,  of $0.14$ at a $0.001$ level.
 %The only insignificant overall effects are those of $R$ and of $S$ on $L$.
 \\[-8mm]

  \section{The dependences for laryngeal cancer on several  factors}
  
  \subsection{Logit regressions for  $\bm L$ as response to $\bm{V,C,R}$}

The observed dependences of laryngeal cancer, $L$,   on heavy vodka drinking, $V$,  are shown first for the two regions and given  regular versus heavy cigarette smoking.\\[-6mm]
   \begin{table}[H]
 \caption{Counts, percentages and observed odd-ratios for  $L,V$ given  $C$ in rural regions}
 \begin{center}
{ \begin{tabular}{l  rrlrr r}
\hline
$L$, laryngeal &\multicolumn{5}{c}{levels of $V,C$ (heavy drinking/ heavy smoking)  }\\
\cline{2-6}
cancer &no/no  & yes/no& \nn &no/yes & yes/yes\\
\hline
controls & 73    &  7 &&26 &2   \\
cases &  (20\%) 18& (42\%) 5&& (24\%) 8&(89\%) 17\\
\hline
sum & 91  & 12 && 34 & 19\\
\hline
${\rm odr}(LV|C, R=0)$& \multicolumn{2}{c}{2.9}&& \multicolumn{2}{c}{27.6} \\
 \hline \end{tabular}}
\end{center}
\vspace{-5mm}
\end{table}

Table 5 is for rural communities. It shows a much stronger dependence of laryngeal cancer on heavy vodka drinking
when the men are heavy cigarette smokers than when they are regular but not so heavy smokers.
By contrast for urban  communities, Table 6 shows a much stronger dependence of laryngeal cancer on heavy vodka drinking
when the men are regular cigarette smokers than when they are heavy smokers. \\[-8mm]

  \begin{table}[H]
 \caption{Counts, percentages and observed odd-ratios for  $L,V$ given $ C$ in urban regions}
 \begin{center}
{ \begin{tabular}{l  rrlrr r}
\hline
$L$, laryngeal &\multicolumn{5}{c}{levels of $V,C$ (heavy drinking/ heavy smoking)  }\\
\cline{2-6}
cancer &no/no  & yes/no& \nn &no/yes & yes/yes\\
\hline
controls & 198    &  9 &&52  &9  \\
cases &  (23\%) 60& (83\%) 43&& (37\%) 30&(72\%) 23\\
\hline
sum & 258 & 52 && 82 & 32\\
\hline
${\rm odr}(LV|C, R=1)$& \multicolumn{2}{c}{15.8}&& \multicolumn{2}{c}{4.4} \\
 \hline \end{tabular}}
\end{center}
\vspace{-5mm}
\end{table}

 The corresponding logit model
is written as $L:\n V*C*R\,$  in Wilkinson's  notation for generalized linear models; see e.g. McCullagh and Nelder (1989). It includes with an important  3-factor interaction effect of $V,C,R$ on $L$   also all main effects and all 2-factor effects so that the parameters  are those of the saturated, i.e. the unconstrained,  logit regression and  the maximum-likelihood estimates of the expected counts agree with the observed counts.

 The 3-factor interaction  in the logit regression is estimated in terms of the four 
observed odds-ratios in Tables 5 and 6 as
$$\log(4.43/15.77)-\log(27.62/2.90)= -3.52$$ with an estimated  standard deviation of $1.22$, giving $z_{\rm obs}=-3.52/1.22=-2.9$. 

We denote any estimated highest-order logit term divided by its standard deviation by $z_{\rm obs}$,   as it can roughly be viewed as  a realisation of a
Gaussian random variable that is standardized to have zero mean and unit variance, provided the population term is zero. With a  p-value of $0.002$, the 3-factor interaction is highly significant so that the dependence structure in Tables 5 and 6 cannot be simplified.

 Alternatively, the same judgement is reached with a likelihood-ratio  goodness-of-fit test  for $L:\, (V+C+R)^2$,
 the  logit model with  just the 3-factor interaction term
 of $V,C,R$ on $L$  removed from the saturated model.
 With a chi-square value of 9.2
on 1 degree of freedom (df) and   $\sqrt{9.2}=3.0$, again roughly the value of a standardized Gaussian distribution if the model fits, the agreement with the absolute value of the above reported $z_{\rm obs}=-2.9$ is high, as expected.\\[-8mm]

\subsection{Dependence of  $\bm L$ on $\bm{V,C,R}$ based on structure for cases and controls}

 For controls and for cases, the observed $VCR$ table of counts are contained  within the tables of  counts shown in Tables  5 and 6.  For controls, these counts are well compatible with $VC\ci R$ so that the rates of heavy versus regular cigarette smokers are  estimated by the overall observed percentages of heavy vodka drinking given $C$, for regular smokers with 5.6\% versus 12.4\%  for heavy smokers.  Thus, the risk for heavy vodka drinking at the level of 5.6\%  is more than doubled when one looks at the control group of  heavy cigarette smokers compared to the control group of regular smokers. 
 
 For cases, there is   a different, more complex dependence  structure for $V,C,R$.   For cases in rural communities,
 vodka drinking is strongly increased when they are also  heavy cigarette smokers; from   22\%  to 68\%.  Thus the risk of heavy vodka drinking starts at a level that is considerably higher than  for any of the   control groups and the relative risk of heavy vodka drinking is estimated to be at a factor of about 3 for heavy compared to regular smokers; with 68/22=3.1. 
  For cases in urban communities instead,  heavy vodka drinking  is observed at essentially the same rates  for regular and heavy cigarette smokers with 42\% and 43\%; see Table 7. \\[-8mm]
   \begin{table}[H]
 \caption{Heavy vodka drinking for cases   by cigarette smoking and living area}\n\\[-12mm]
 \begin{center}
{ \begin{tabular}{l  rr  l rr}
\hline\
$V$, heavy&\multicolumn{2}{c}{$R$ rural, $C$ heavy smokers}&\nn &\multicolumn{2}{c}{$R$  urban, $C$ heavy smokers}\\
\cline{2-3} \cline{5-6}
vodka drinking & no& yes & & $ no $  & yes\\
\cline{1-3} \cline {5-6}
0:= no & 18    &  8& & 60 & 30\\
1:=yes &  (22\%) 5& (68\%) 17&& (42\%) 43&(43\%) 23\\
\cline{1-3} \cline{5-6}
${\rm odr}(VC|R)$& \multicolumn{2}{c}{7.7}&& \multicolumn{2}{c}{1.1} \\
 \hline \end{tabular}}
\end{center}
 \vspace{-5mm}
\end{table}

These basic data summaries in terms of  the observed counts point to the differences in exposure that explain already to a large extent the  observed interactive effect of $V,C,R$ on $L$. They supplement the maximum-likelihood estimates of the probabilities at the two levels of $L$ which can be expressed formally for $L=0$ and $L=1$ as
$$ \hat{\pi}^{LVCR}_{0jkl}=n_{0jk+}n_{0++l}/n_{0+++}^2 \text{ and } \hat{\pi}^{LVCR}_{1jkl}=n_{1jkl}/n_{1+++}\,,$$
when $n_{ijkl}$ denotes the observed count in the $LVCR$ table and summing is indicated by the $+$-notation in the same way as for probabilities. These are the estimates for   two graphical log-linear models; see Darroch, Lauritzen and Speed (1980). Their graphs are also known as concentration graphs. The model for the controls has  tables $VC$ and $R$ as the set of  minimal sufficient statistics and the model for the cases needs table $VCR$ to generate the joint distribution.

These types of models have  representations in terms of  undirected graphs with nodes for variables but at most one full-line edge coupling each node pair. For exclusively categorical variables, the set
of minimal sufficient statistics of the model coincides with  the set of cliques in the graph. For a clique, each node is coupled to all other nodes within the clique, and an incomplete graph  is induced when just one further node of the graph is added to the nodes of a clique.
In  case-control studies these models arise as  independence structures that one can study directly for cases and for controls separately.

No graph distinguishes between additive and interactive effects, but each graph permits to read off  the graph all independence statements that the model implies; the criterion for full-line graphs is given in connection with  Figure 3 below. The two graphs  relevant here for $VCR$ given $L$ are the two simple graphs in Figure 2.

For the cases, there is a complete graph so that no independences are implied. For the controls, nodes $VC$ are coupled by an edge and are isolated from  node $R$. This captures
$VC\ci R$  as well as independences  that follow from the model, for instance, $V\ci  R|C$ or $V\ci R$. 
\begin{figure}[H]
\centering
\includegraphics[scale=.52]{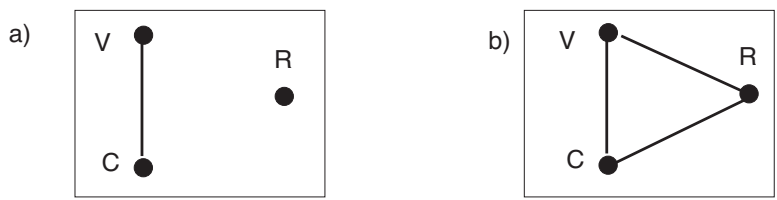}
\caption{\small Graphs capturing the independence structure of $V,C, R$ a) for controls, b) for cases.}
\label{figlabin}
\end{figure}

By combining the counts estimated separately for controls and cases, shown in Table 8,  
smoothed  odds-ratios result compared to a  logistic regression which uses implicitly the structure of
two complete concentration graphs. The graphs summarize the estimated independence structure  in the populations of either the cases or of  the controls, for  a  set of  variables that is directly explanatory for the disease.

  \begin{table}[H]
 \caption{Estimates, obtained with separate models of $VCR$ for cases and controls}
 \begin{center}
{ \begin{tabular}{ll  rrlrr l rrlrr}
\hline
$L$, laryngeal &&\multicolumn{11}{c}{levels of $V$, heavy vodka drinking; $C$, heavy  smoking; $R$, rural}\\
\cline{3-13}
cancer && \n 000 & \n 100 &\nn& \n  010 &\n 110&\nn   \nn &  \nn  001& \n 101&\nn& \n 011& \n111\\
\hline
controls &&   \n 77.8 &        4.6   &  &   22.4&         3.2&&      193.2& 11.4& &      55.6&         7.8\\
cases &  & \nn18&    5  &&    8  &  17  &&  60 &   43  && 30  &  23\\
\hline
$ \hat{\odds}{\rm dr}(LV|CR)$&& \multicolumn{2}{c}{\nn 4.7}&&\multicolumn{2}{c}{\nn15.1}&& \multicolumn{2}{c}{\nn 12.1}&&\multicolumn{2}{c}{\nn5.4}\\
 \hline \end{tabular}}
\end{center}
\vspace{-5mm}
\end{table}
Smoothed, estimated odds-ratios obtained from counts in several cliques of graphs  are based  on observations for more individuals than those using complete graphs. This leads  to smaller standard deviations of the estimated odds-ratios and hence to more reliable  results;  provided only that the structures in the  populations are well estimated.

The larger the  numbers of observation, the more likely it is that this goal is reached, unless there are 
variables that show extremely strong dependences.  Even if  results appear to be reliable in this
sense, 
confounding effects of additional variables need to  be ruled out, here those of $A,E,S$. In the next sections, we  do this first by using logistic regressions with  enlarged sets of regressors,
next by exploiting again independences.

\subsection{Effects of $\bm A$, $\bm E$  or $\bm S$ on $\bm L$ in logit regressions when added to $\bm{V,C,R}$}
 To contemplate directions of effects of $A$, of  level of formal schooling, $E$,  and of age when starting to smoke, $C$, is to follow R.A. Fisher's
advice of making `theories elaborate' as quoted by Cochran (1965), that is
`when constructing a causal hypothesis one should envisage as many different consequences of its truth as possible'.

Since age at the time of study, $A$, is for cases partly a proxy for  cumulated exposure to several risk factors, currently known or still unknown, one can also expect that the illness  occurs more frequently the
higher the age at study time, so that there is an increasing, additive  effect of $A$ on $L$.
The level of formal schooling, $E$, could have a similar effect since it correlates strongly with $A$. On the other hand, since there is no overall effect of $S$, age when cigarette smoking was started, one expects no direct effect of $S$, when $V,C,R, A$ or $V,C,R, E$ are already 
used to predict $L$. 

 In the logit models for $L$ that include in addition either   age at study time, $A$,  or level of education, $E$,
the three-factor effect term for $VCR$ remains significant and there is an additional significant main effect of the added variable. The goodness-of-fit of each of the models can  then be tested and is very good for model $L:\, V*C*R+A$
with a chi-square value  of  $4.1$ on 7 df, it is a bit less convincing,  looked at alone,  for model $L:\, V*C*R+E$, where there is  a chi-square-value value of $10.1$ on 7 df. %More details  are given in Appendix 2.

Since the observed tables of counts for $LVCRA$ and $LVCRE$ are too sparse to report sensible   percentages, Table 9 shows 
  only the observed odds-ratios. They confirm that the general pattern of the dependences in Tables 5, 6 is approximately repeated given  the two levels of  either $A$ or of $E$, so that an additive main effect, in addition to the interactive effect of $VCR$  is a  plausible description
of the observed odds-ratios. Indeed, they are close to
the odds-ratios estimated under the two models  $L:\, V*C*R+A$ or  $L:\, V*C*R+E$, shown in Table 10.\\[-10mm]%The table $VACRL$ of counts and estimated percentages is given in the next section.
 \begin{table}[H]
 \caption{Observed odds-ratios for  laryngeal cancer, $L$, and heavy vodka drinking, $V$}
 \begin{center}
{ \begin{tabular}{l  rrrrrrrr}
\hline
&\multicolumn{8}{c}{level combinations of $CRA$ or $CRE$}\\
\cline{2-9}
   & 000  & 100& 010& 110   & 001  & 101& 011& 111 \\ \hline
odr$(LV|CRA)$ &  1.4    &     34.0 &        11.2 &          3.5 &          5.3 &        19.8&         20.3&          4.2\\
odr$(LV|CRE)$&   -      &         27.0&         18.5 &         15.5&          8.0&        34.0&         12.3&         2.6\\
\hline
 \end{tabular}}
\end{center}
\vspace{-5mm}
\end{table}

In  Table 10, the estimates at level 0  of $A$ are repeated at level 1 of $A$.  
At all levels of $CRA$ or $CRE$,  a  strong dependence of laryngeal cancer  on heavy vodka drinking is estimated, for instance
 with an odds-ratio of  $ \hat{\odds}{\rm dr}(LV|C=0, R=0, A=0)= 3.1$,    for rural regular smokers in the younger age group and with $ \hat{\odds}{\rm dr}(LV|C=1, R=0, A=0)= 27.3$
  for rural heavy smokers in the older age group.

 \begin{table}[H]
 \caption{Estimated odds-ratios for  $L$ and $V$
 for models  $L\,: VCRA$ and  $L\,: VCRE$}
 \begin{center}
{ \begin{tabular}{l  rrrrrrrr}
\hline
&\multicolumn{8}{c}{level combinations of $CRA$ or $CRE$}\\
\cline{2-9}
             & 000  & 100  & 010  & 110 & 001 &  101 & 011  & 111 \\ \hline
$ \hat{\odds}{\rm dr}(LV|CRA)$&  3.1 & 27.3 & 14.4 & 3.8 & 3.1 & 27.3 & 14.4 & 3.8\\
$ \hat{\odds}{\rm dr}(LV|CRE)$&  3.6 & 30.9 & 14.3 & 4.5 & 3.6 & 30.9 & 14.3 & 4.5\\
\hline
 \end{tabular}}
 \end{center}
 \vspace{-5mm}
\end{table}

Instead variable $S$, age when cigarette  smoking had been started, does not improve prediction of $L$ if added to $V,C,R,A$ since none of the one df components is large for  15.4 on 16 df,  the chi-square value resulting  when model   $L \ci S|VCRA$ is fitted.   Overall, these added logit regressions appear to support the hypotheses about the effect directions. However,  more insight is gained by  exploiting again independences present for cases or for controls.\\[-8mm]

 \subsection{The independence models $\bm{A\ci VR|CL}$ and $\bm{V\ci R|ACL}$}

 For $VACR$ given $L$, we  compare first   two independence models that are graphical log-linear models,   $A\ci VR|CL$ and $A\ci V\ci R|L$, with the  logit regression model $L: A+V*C*R$. 
Several special features show by conditioning  explicitly on cases and controls.

The graph of  model $A\ci VR|CL$  is given in Figure 3. 
For every   statement  $a\ci b|c$ implied by  this model for disjoint subsets $a,b,c$ of $\{A,V,R,C,L\}$,  every path from $a$ to $b$ has a node in $c$. Thus in particular, every path from $A$ to $\{V,R\}$  passes through $C$ or $L$ and, for instance with strong dependences of $L$ on each of  $V,A$, a  dependence between $V$ and $A$, which is absent given $L,C,R$, will be introduced by marginalising over $L$. \\[-5mm]
\begin{figure}[H]
\centering
\includegraphics[scale=.51]{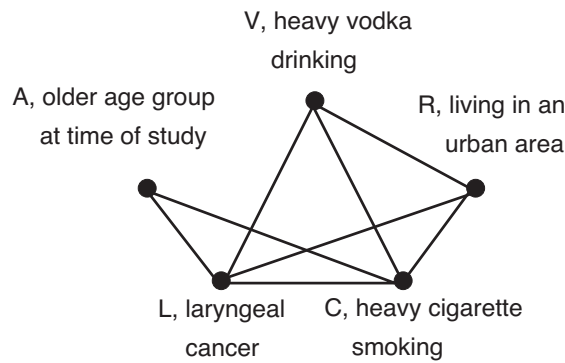}\n\\[-1mm]
\caption{\small Concentration graph of model $A\ci VR|CL$; likelihood chi-square: 4.7 on 12 df.}
%with count estimates given in Table 11.}
\label{labinconc}
\end{figure}

Table 11 shows how little the observed 
counts deviate from  estimates under model  $A\ci VR|CL$ that is  for heavy vodka drinking, $V$, and 
urban  living area, $R$, conditionally independent of the age groups at study time, $A$, given $C$  and    cases or controls. It also contains the fitted counts for model $A\ci V\ci R|L$ and the 
 $VARC$ table.

\begin{table}[H]
  \caption{Observed  counts for the $VARCL$ table and estimates for $A\ci VR|CL$} \n \\[-12mm]  
   \begin{center}
\begin{tabular}{ccccrrrr rrr  rr}
\hline
& &&& \multicolumn{3}{c}{controls} &&\multicolumn{3}{c}{cases} &\multicolumn{2}{c}{combined}\\
 \cline{5-7} \cline{9-11}
\multicolumn{4}{c}{$VACR$ levels}&count& estim.$\!^{*}$&perc.$\!^{*}$&&count& estim.$\!^{*}$&perc.$\!^{*}$& count&percent\\
\hline
     0  &   0  &   0  &   0  &     41    &     43.24      &           & & 6    &      4.14    &      &  47 &   \\
     1  &   0  &   0  &   0  &       5    &      4.15     &      8.7   && 1    &      1.15    &     21.7&   6 &   11.3 \\
     0  &   1  &   0  &   0  &      32    &     29.76          &     & &12    &     13.86    &      &  44 &      \\
     1  &   1  &   0  &   0  &       2    &      2.85  &      8.7   &  &  4    &      3.85    &     21.7 &   6 &      12.0\\
     0  &   0  &   1  &   0  &      17    &     16.94        &      &&    3    &      3.08    &      & 20 &      \\
     1  &   0  &   1  &   0  &       1    &      1.30    &     7.1  & &   6    &      6.54    &     68.0& 7 &      25.9\\ 
     0  &   1  &   1  &   0  &       9    &      9.06           &     & &   5    &      4.92    &      & 14 &       \\ 
     1  &   1  &   1  &   0  &       1    &      0.70    &      7.1 &&   11    &     10.46    &     68.0 &  12 &      46.2 \\
     0  &   0  &   0  &   1  &     118    &    117.28       &    & &   14    &     13.81    &      &  132 &      \\
     1  &   0  &   0  &   1  &       6    &      5.33    &      4.3  & &   8    &      9.90    &     41.7  &   14 &       9.6\\
     0  &   1  &   0  &   1  &      80    &     80.72        &      &&   46    &     46.19    &     &126 &      \\ 
     1  &   1  &   0  &   1  &      3    &      3.67   &      4.3 &    & 35    &     33.10    &     41.7&  38 &      23.2\\
     0  &   0  &   1  &   1  &      35    &     33.89           &     & &14    &     11.54    &     &  49 &       \\
     1  &   0  &   1  &   1  &       5    &      5.87        &     14.8 & & 7&         8.85    &     43.4 &  12 &      19.7\\
     0  &   1  &   1  &   1  &     17    &     18.11         &      &   &16    &     18.46    &      &  33 &      \\ 
     1  &   1  &   1  &   1  &       4    &      3.13    &     14.8 &   &16    &    14.15     &  43.4 &  20 &      37.7\\ 
  \hline
\end{tabular}
\end{center}\n \\[-6mm]
\small $^{*}$estimated counts for model $A\ci VR | CL$ with deviance 4.7 on  12 df; percent for estim. $V=1$

\end{table}

 By  mixing the  samples from  the two different populations, the cases and the controls, for the $VARC$ table, a sizeable  dependence of  $V$ on $A$ given $RC$ results.  This shows in the  percentages in the last column of Table 11,  which   vary widely within some of the fixed levels of $C,R$. 
If we had data from  a prospective study with  $L$ as   response  occurring in the future of the explanatory variables, it would be impossible to induce any dependence by marginalizing over $L$. 

For interpretation, the important feature of model $A\ci VR|CL$ is that it implies  $V\ci A|CRL$.  
Therefore, for heavy vodka drinking, $V$, marginalizing over $A$ leaves the dependence of $V$ on  $C$ unchanged for all level combinations of $R,L$. This explains why the estimated rates of heavy vodka
drinking given this model coincide with those observed for the $VCR$ tables at both levels of $L$. In particular, the estimated rates for cases in Table 11, repeat for the two levels of $A$ and are identical
except for rounding errors to the rates for cases in the observed $VRC$ table of Table 7.

For model $V \ci R|CVL$, the fit seems to be good with no large component chi-square value in the eight tests for independence of pair $V,R$ defined by  the combinations of $C,V,L$; the sum giving a chi-square value of 11.0 on 8 df and a  p-value of 0.20. Its graph arises from the one in Figure 3, by adding the two edges $A  \ful V$, $A \ful R$ and removing $V\ful R$. However, the fitted odds-ratios for $L,V$ differ substantially from
the observed values in Table 9, while those of the other  two models in Table 12 agree well.

\begin{table}[H]
 \caption{Odds-ratio estimates for  $L, V$  given $CRA$ under selected  models}\n\\[-12mm]
 \begin{center}
{ \begin{tabular}{r  rrrrrrrr}
\hline
&\multicolumn{8}{c}{level combinations of $CRA$}\\
\cline{2-9}
 $ \hat{\odds}{\rm dr}(LV|CRA)$ for:            & 000  & 100  & 010  & 110 & 001 &  101 & 011  & 111 \\ \hline
logit \nn   \nn   \nn  $L:\, A+V*C*R$&  3.1 & 27.3 & 14.4 & 3.8 & 3.1 & 27.3 & 14.4 & 3.8\\[2mm]

log-linear\nn  \nn \nn \nn \,$A\ci VR| CL$&         2.9 &        27.6  &       15.8   &       4.4   &       2.9  &       27.6 &        15.8    &      4.4 \\
$V\ci R|ACL$&          6.5 &         6.6  &        6.5   &       6.6   &      15.1  &        6.7&        15.1    &      6.7  \\
\hline
 \end{tabular}}
 \end{center}
 \vspace{-5mm}
\end{table}

This can be explained. The sparser a table gets, the more difficult it becomes to decide among seemingly  
well-fitting models.  If here  the independence of pair $V,R$ is considered in addition to the extremely well-fitting model $A\ci V|RCL$ having a p-value of 0.97,
then it becomes a test in the marginal table $VCRL$ and leads to a poorly fitting model with a p-value of only 0.03; derived from  a chi-square value of 10.7 on 4 df. 
\\[-8mm]

 \subsection{Further mixed count estimates to obtain smoothed odds-ratios}
 Though there is with $A\ci VR|CL$ a well-fitting independence structure for both cases and controls, we expect substantial differences of
 the dependences and independences in the two populations. After all,  the main purpose of a case-control study is to gain an understanding
 of what the differing features and dependences are of those  who have been diagnosed to have the disease from  those who have not.

 For cases and  controls analyzed separately, two well-fitting structures 
are captured by the graphs in  Figure 4. They may for instance be derived by  forward selection in the general class of concentration graphs with a goodness-of-fit criterion for the selected edges of $p=0.2$; see H\o jsgaard, Edwards and Lauritzen (2012).\\[-4mm]

\begin{figure}[H]
\centering
\includegraphics[scale=.51]{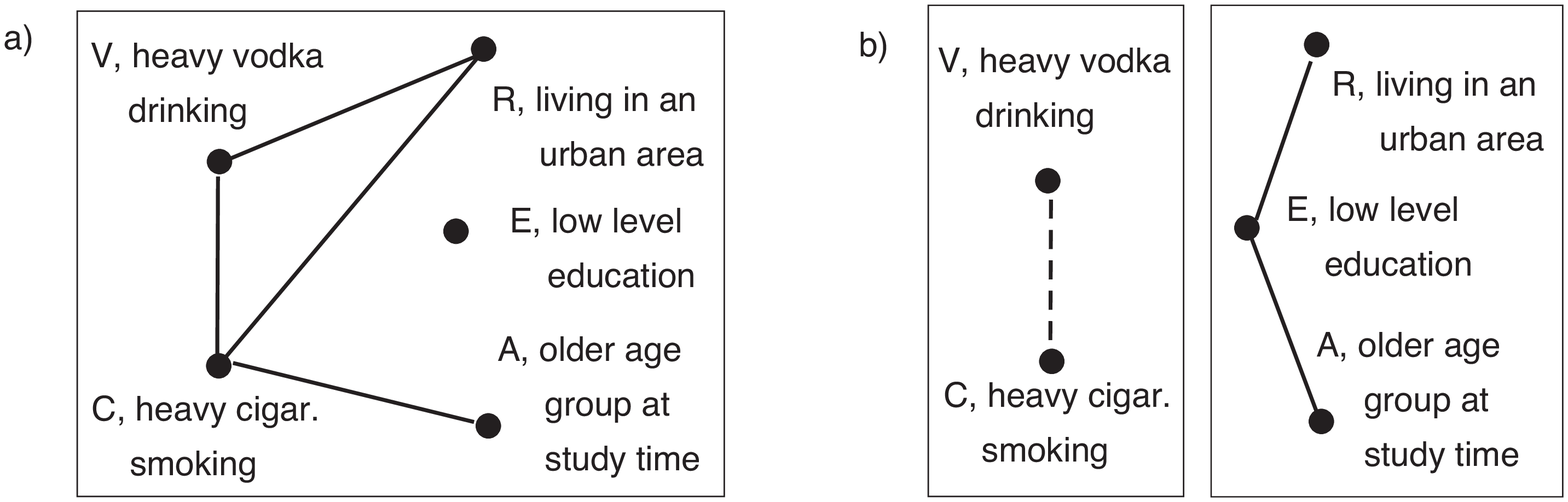}\n\\[-4mm]
\caption{\small Graphs of two separate  well-fitting models  for $V,C,R,A,E$
a) a concentration graph for cases with a fit of 16.2 on 21 df; b) a regression graph for controls with 17.9 on 23 df.}
\end{figure}

 The count estimates are reported in Table 19 in the Appendix and the estimates of the conditional odds-ratios based on the separate models for cases and controls to the graphs in Figure 4 are shown in Table 14 below. 
By marginalizing in both graphs of Figure 4 over $E$ and $A$, no additional edge is induced; see Wermuth (2011). The two concentration graphs in Figure 2 arise, the first four smoothed estimates of the odds-ratios in row 2 of Table 14  coincide  with those estimated  in Table 8 from the  $LVCR$ table alone, and they are repeated in unchanged form for all level combinations of $A,E$.

 For cases, Figure 4a) shows  the selected  concentration graph for which the  overall fit of the model  to is given  by a chi-square value of 16.2 on 21 df. 
  For the controls,  Figure 4b) shows the selected regression graph which is Markov equivalent to a full-line concentration graph in the same nodes and in the same set of edges, since there is no collision {\sf V} in the regression graph. The overall fit of the model  in Figure 4b) is given  by  a  chi-square value of 17.9 on 23 df; for general fitting procedures of regression 
 graph models for categorical variables; see Marchetti and Lupparelli (2011).
              
 The  overall  goodness-of-fit  of these graphical log-linear   models can typically  be decomposed into  a sequence of much smaller fitting steps.  For  instance for cases, the model
 $E\ci VCRA$ splits   into the sequence $E\ci A|VCR$, $E\ci V|CR$, $E\ci V|R$, $E\ci R$
 with chi-square values of  $ 5.3, \n 3.7,  \n   2.9          \n 1.2$   on $8, \n 4, \n 2, \n 1$ df, respectively, while for
the additional fit of  $A\ci VR|C$  with 3.0 on 6 df,  no single degree of freedom component  can be significant.
          
Table 13 contains the estimates in two logit regressions. Choosing as  highest order interactions  terms the cliques in Figures 4a) or 4b) assures a good fit. 
\\[-10mm]
\begin{table}[H]
\begin{center}
\caption{Estimates for  logit models of laryngeal cancer as response}\n \\[-3mm]\label{fit3}
\begin{tabular}{lrrrcrrr} \hline
response $L$ & coeff    &  $s_{\rm coeff}$   &  $z_{\text{obs}}$    &  &   coeff & $s_{\rm coeff}$& $z_{\text{obs}}$  \\ \hline
(const.)     & $-3.49$  & $0.63  $ &     ---  &  & $-3.37$  & 0.42     & ---      \\
$V$          & $1.33$  & $0.72$   & --- &  & $ 1.32$  & $0.71  $   &  --- \\
$C$          & $0.64$  &  $0.57$  &  ---     &  & $0.29$  & 0.50     & ---     \\
$R$          & $0.30$  &$ 0.64$   &  ---     &  & $0.34 $  & 0.32     & ---     \\
$VC$        & $2.04$  & 1.14  &  --- &  & $2.08$  &$ 0.16   $  & --- \\ 
$VR$        & $ 1.31$&$0.84$&---&     &    $1.30$&     $0.83$    &---       \\
$CR$        & $0.50$&$0.58$&---&   &      $  0.50$& $0.58$ &---              \\
$VCR$      &  $-3.34$& $1.32$& $-2.53$    & & $-3.39$& $1.32$& $-2.56$  \\
$A$            & $2.56$  & $0.46$   &  ---     &  & $2.36 $  &  0.43   & ---            \\
$CA$            & $-0.59$  & $0.46$&$-1.26$   &    & --- & --- & ---             \\
$E$             &  $ 1.95 $ &$0.64 $ &  ---     &  & $1.96$  &  $0.38$   & ---            \\
$AE$          &  $ -1.88 $ &$0.50 $ & $-3.77$      &  & $-1.87$  & $ 0.49$   &  $-3.79$         \\
$ER$           & $0.04$& $0.65$&$ 0.06$ &  &   ---&---& ---              
\\ \hline
 \end{tabular}
  \vspace{-5mm}
\end{center}
\end{table}
 Wilkinson's notation for the  first  logit models is $L:\, V*C*R+C*A+A*E+E*R$. This logit  model  is  equivalent to the log-linear model which has as  minimal sufficient tables, $LVRC,LCA, LAE, LAR$ and $VCRAE$, where the last  table is the one  of the regressor variables; see Haberman (1974). The fit is given by a chi-square value of $19.8$ on  19 df. 

The model can be simplified by excluding just two non-significant interaction terms $CA$ and $ER$. This leads to model $L:\, V*C*R+A*E$ with a fit given by 21.4 on 21 df.  The additivity of the effects of $V,C,R$ and of $A,E$   in  this logit regression model explain  why the odds-ratios for instance of $L,V$,
estimated under model  $L:\, V*C*R+A*E$, repeat in unchanged form for all level combinations of $A,E$.

\begin{table}[H]
 \caption{Odds-ratio estimates for  $L, V$  given $CRAE$ under  two selected  models, shown
 only for $E=0$;  estimates are repeated in unchanged form for the eight levels of $E=1$}\n\\[-17mm]
 \begin{center}
{ \begin{tabular}{r  rrrrrrrr}
\hline
&\multicolumn{8}{c}{level combinations of $CRA$ given $E=0$ }\\
\cline{2-9}
 $ \hat{\odds}{\rm dr}(LV|CRA)$ for:         & \nn 000  & 100  & 010  & 110 & 001 &  101 & 011  & 111 \\ \hline
logit model  \n $L:\, A*E+V*C*R$& 3.8  & 30.1 & 13.7  & 3.7 & 3.8 & 30.1 & 13.7  & 3.7\\[2mm]
count estimates to Figures 4a), 4b)  &        4.7 &        15.1  &       12.1   &       5.4   &       4.7  &       15.1 &        12.1    &  5.4 \\
\hline
 \end{tabular}}
 \end{center}
 \vspace{-5mm}
\end{table}
As before in Table 8, the estimates based of the separate well-fitting concentration graphs for cases and controls are more smoothed than those of the minimally fitting logit  regression. From  both approaches, one concludes that each of  five regressor variables is an important explanatory
variable but that age at study time and level of education do not confound the dependence of $L$ on $V$
given $C, R$.

\begin{figure}[H]
\centering
\includegraphics[scale=.51]{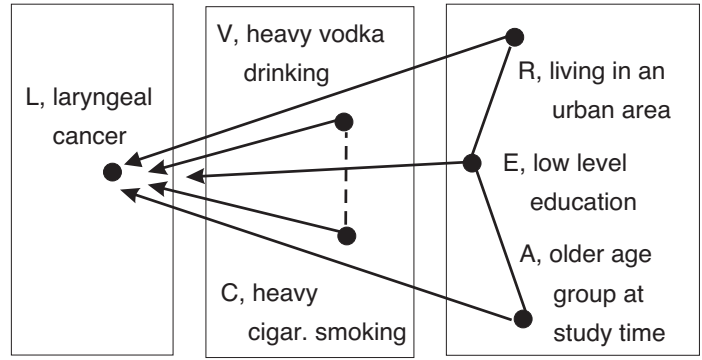}\n\\[-4mm]
\caption{\small A regression graph summarizing important dependences; for $L$, it shows the relevant explanatory variables, but for $V,C,R,A,E$  the independence structure among controls only.}
\end{figure}
                            
For the general population, the fitted regression graph in Figure 5 and Figure 4b) imply in particular  $V\ci AE|CR$ since  this follows from the  complete  independence  $VC\ci REA$. Thus  by the collapsibility criterion of equation (5),   the   constant increase of the  conditional  relative risks of laryngeal cancer with  heavy  vodka drinking for  both groups of smokers in rural and urban communities do not change  with  the two age groups or with  the two levels of formal education.   

The differences in the  cliques of  the two concentration graphs  for $V,C,R,A,E$  to Figure 4 point instead  to how cases differ mainly from controls with respect to the  relations among these five features.\\[-8mm]

 \section{Understanding differences between cases and controls}

 Observed and estimated counts that exploits structure in the two separate samples of cases and controls  are given for the $VRCAEL$ table in Table 19 of the Appendix. We use them now together with the   cliques
 in the two graphs of Figure 4  to appreciate  how the  cases differ from the  controls.

 The two graphs in Figure 4 have three main distinguishing features. The subgraph induced by nodes $VCR$, 
 is  complete in Figure 4a) and a  $VC$-edge isolated from node $R$ in Figure 4b).  The corresponding
 differences in exposure to heavy vodka drinking and cigarette smoking have already been
 described in the discussion of the observed $VCR$ table in Section 3.1.
 Then, there  is a {\sf V} with inner node $E$ connecting 
 nodes $A$, $R$ in Figure 4b) and no such path in 4a). Finally, there is an $AC$-edge  in Figure 4a) and no such edge in 4b). 
 
   For controls, the graph in Figure 4b)  implies $A\ci R|E$, so that the almost equal observed  odr$(EA|R=0)=7.2$ and odr$(EA|R=1)=7.5$ are, by equation (4), collapsible to a value near odr$(EA)=7.1$. Similarly,   the conditional  odds-ratios for $E,R$ given $A$ reduce by equation (4) to odr$(ER)=0.5$; see also Table 15. \\[-8mm]

  \begin{table}[H]
 \caption{Lower level of education  for controls by age and by living area}\n\\[-12mm]
 \begin{center}
{ \begin{tabular}{l  rr  l rr}
\hline\
$E$, low&\multicolumn{2}{c}{$A$, age group at study time}&\nn &\multicolumn{2}{c}{$R$, living area}\\
\cline{2-3} \cline{5-6}
level educ. &33--50  & 51--65& & urban &rural\\
\cline{1-3} \cline {5-6}
0:= no & 151    &   32& & 40 & 143\\
1:=yes &  (34\%) 77& (78\%) 116&& (62\%) 68&(47\%) 125\\
\cline{1-3} \cline{5-6}
sum & 228 & 148&&  108& 268\\
 \hline \end{tabular}}
\end{center}
 \vspace{-5mm}
\end{table}
 
 For cases instead, the graph in Figure 3a) implies $E\ci AR$. The observed percentages of cases with a low educational level
 are essentially  the same level for both age  groups in both living areas and estimated  by the high overall rate of 
 almost 80\%  (162 of 204). Thus, there is a kind of ceiling effect.  Since almost all cases have the same low educational level, no effect of education can show up for the cases.

 For both controls and for cases, a smaller percentage of the men in the older age group are heavy smokers; see Table 16. But the change from
 25\% to 21\% for controls is not large enough  to be statistically significant. For cases, the decrease is  from 51\%
 to 33\%; significant at a 0.05 level with 5.5 on 1 df. 
 \\[-8mm]
   \begin{table}[H]
 \caption{Heavy cigarette smoking for cases and for controls  by age}\n\\[-12mm]
 \begin{center}
{ \begin{tabular}{l  rr  l rr}
\hline\
$C$, heavy&\multicolumn{2}{c}{$A$, age group; controls}&\nn &\multicolumn{2}{c}{$A$, age group; cases}\\
\cline{2-3} \cline{5-6}
cigar. smoking &33--50  & 51--65& & 33--50  & 51--65\\
\cline{1-3} \cline {5-6}
0:= no & 170    &   117& & 29 & 97\\
1:=yes &  (25\%) 58& (21\%) 31&& (51\%) 30&(33\%) 48\\
\cline{1-3} \cline{5-6}
sum & 228 & 148&&  59& 145\\
 \hline \end{tabular}}
\end{center}
 \vspace{-5mm}
\end{table}

As shown in Section 3, the risks for laryngeal cancer  are highest  for rural heavy smokers and for urban regular smokers;
the features of these two  groups are summarized next. \\[-8mm]
    \begin{table}[H]
 \caption{Comparing features of cases with controls for urban, regular  smokers}\n\\[-12mm]
 \begin{center}
{ \begin{tabular}{l  rrrrrrrr}
\hline
$L$, laryngeal &\multicolumn{3}{c}{Observed percentages; counts}& \n&\multicolumn{3}{c}{Means; (st.dev.)}\\
\cline{2-4} \cline{6-8}
cancer & $V=1$  & $A=1$& $E=1$&& $cV\;\!^{*}$& \n$cA\;\!^{*}$& $cC\;\!^{*}$\\
\hline
controls; $n=207$ &  (4\%) \n9  &  (40\%) 83  &(48\%) 99  && 1.1& 46.8&17.2 \\
&&&&& (5.3)& (10.1) & (3.2)\\
cases; $n=103$& (42\%) 43&  (79\%) 81 & (75\%)  77 && 11.1& 54.1 & 18.6 \\
&&&&& (13.9)& (6.4) & (2.2)\\
 \hline
 \multicolumn{6}{l}{\small{where e.g. $cA\;\!^{*}$ means that age is viewed as being continuous feature}}
 \end{tabular}} \end{center}
\vspace{-5mm}
\end{table}

In both Tables 17 and 18, one sees when changing from controls to cases,   a tenfold increase in the rate of heavy vodka-drinking, 
a doubling of the risk of being in the older age  group and a clear increase in the risk of having a lower
level of education.

                   \begin{table}[H]
 \caption{Comparing features of cases with controls for rural, heavy  smokers}\n\\[-12mm]
 \begin{center}
{ \begin{tabular}{l  rrrrrrrrr}
\hline
$L$, laryngeal &\multicolumn{3}{c}{Observed percentages; counts}& \n&\multicolumn{3}{c}{Means; (st.dev.)}\\
\cline{2-4} \cline{6-8}
cancer & $V=1$  & $A=1$& $E=1$&& $cV$& \n$cA$& $cC$\\
\hline
controls; $n=28$ &  (7\%) \n\,2  &  (36\%) 10  &(68\%) 18 && 0.9&47.7&30.8  \\
&&&&& (3.7)& (10.1) & (8.8)\\
cases; $n=25$& (68\%) 17 &  (64\%) 16 & (84\%) 21 && 14.7 &51.7& 29.1\\
&&&&& (11.4)& (6.7) & (7.8)\\
 \hline \end{tabular}
}
\end{center}
\vspace{-5mm}
\end{table}

 Even though the groups in Tables 17 and 18 differ by design in the amount of tobacco consumption,
 the  relative risks of  of laryngeal cancer due to heavy vodka drinking 
are similar when estimated as 15.1 and 12.1  using the separate count estimates for cases and controls;
see Table 14.  
In particular  cases and controls are, within both groups, quite comparable regarding the average number of cigarettes  smoked per day. 
But within both living areas,  the men who are cases
are on average  older than those who are controls, a much larger proportion of the cases has a low level of formal  education compared to the controls and heavy vodka consumption increases from 
an average of just one year of exposure to more than 10 years. Thus,  these negative  three  features accumulate in  a similar way in the two highest-risk groups of cases.

Nevertheless, it remains unexpected that no similar increase in risk is estimated for the urban communities   as for  the rural ones, when  the 
tobacco consumption is increased  to more than a pack of cigarettes per day.\\[-8mm]

 \section{Interpretation and discussion}

 \subsection{General   issues in the current case-control study}
 
  One main  possibility to explain the differences  in the two areas of  living is that there may have been other risk factors at work in rural and in urban Polish communities at the time of study. These unmeasured competing risks may for instance be  nutritional,  environmental or occupational.   If for instance, there was a higher average morbidity of men in urban communities, more of those who have been heavy smokers and vodka drinkers may have died from other causes before they could have participated in the study.

More specifically, the study group consists of men aged between 33 and 65 years in the second half of 1980, thus they all suffered as children or young adults from
 poor living conditions during and after the second world war.
  Since   the nutritional status was at the time in most European countries on average much worse in  urban than in rural communities, this may have led to a higher level of morbidity in urban than the rural participants.

 In addition, in Polish cities the heating was frequently centralized,
using coal-fired power stations nearby, while wood was preferably  used for heating in rural areas.  Thus  exposure to coal dust or to  the  generally higher  level of air pollution in the urban than in rural communities are plausible differences not reflected in the available data.

Drinking vodka of a particular poor quality or  self-produced vodka, may be further risks that 
differ by regions.  All these are hypotheses that need to be  confirmed or rejected using more information on the participants of the present study or in similar studies of  risks for laryngeal cancer.

Distortions  may also have been introduced by the way in which the study participants were selected or in which they responded in the interviews.  It has has been appreciated a long time ago, that whenever there is individual case-control matching on some variables, one gives up the chance of studying an interactive  effect of the matching variables; see McKinlay (1977), but this  argument does not apply to the  sampling in the current study.
Another  source of possible distortions is  a differential exposure-specific consent in controls and cases.

There was a  high response rate for the controls while the case consent rate was  lower.    In particular, because of the long-term exposure of the  cases to the risk factors, some of the men may  have been too ill to respond to the extended interview. This is a problem common to most case-control studies concerning a  life-threatening illness.

However,  an accumulation of  several negative feature was seen in the current study for persons at highest risk.  When similar analyses of further case-control data  confirm these special combinations, it could help to set up 
feature-specific screening programs or
help physicians to decide when to check for direct early symptoms in their own patients.
\\[-8mm]

\subsection{A summary of the proposed strategy of analysis.}
 
 The  proposed new approach to analyzing case-control data aims at supplementing
 inference based modeling with evidence based arguments and to exploit the special features
 of case-control sampling for improved estimation of risks and for understanding major differences 
 between cases and controls.
 
 After the main
research questions are set out, one defines categorical variables that lead to  a cross-classification with at most a reasonable number of non-empty cells, to correlations that do not differ substantially from those of the raw data and to subgroups  of cases and controls that are comparable with respect to important features known to relate to the disease under study. 

The special sampling scheme of case-control studies  always leads to samples from
two different populations,  one of the cases and one of the controls. After characterizing  the independences and dependence structures supported by  the separate samples for the  regressor variables, the cliques of corresponding  graphs lead directly to  a well-fitting logit regression model. 
 However, in such a logit regression, effects that are statistically significant only  in the sample of the 
controls or of the cases but not in both, need not show as being important and hence could remain undetected.

Therefore,  to identify  important differences between cases and controls,   the  two separate well-fitting graphs, for the  regressor variables alone,  are combined with the relevant observed counts and basic data summaries. This leads to direct evidence which of the observed features of the diseased accumulate
in a different way than in the general population from which the controls are sampled.

The goodness-of-fit tests used  in the logit regression with the binary disease as response and
for  finding differential structure among the regressor variables alone depend  on the saturated models obtained  after  the initial data processing steps. Thus,
there is a remaining danger of  relying too much on goodness of fit tests with  these saturated
models as reference.   But, re-analyses of larger  case-control studies are possible and feasible.
They may help to confirm the more tentative results of a smaller study.

With more data than analyzed here,  alternative data processing steps can also be contemplated which use more refined stratifications than binary variables can offer.  The separate  dependence structures of cases and controls together convincing data summaries to supplement  model based estimates, are  the  key for gaining more insights  than with logistic regressions alone.

   \subsection* {Appendix: Mixed count estimates  for the two  $\bm{VCRAE}$ tables} 
     From the  observed counts for $VCRAEL$ and the estimated counts, obtained separately
   for cases and controls using the well-fitting models to  Figures 4a) and 4b),
   the odds-ratios in Table    14 are obtained. For instance for the observation $21, 2, 4, 0$, in the first $LV$  table at levels zero of each of the four remaining variables, the 
   estimated counts,  $23.79,  0.85  , 1.40, 0.24,$ lead to an estimated  odds-ratio   of $4.7$. 
      
\begin{table}[H]
  \caption{Observed  and estimated\;$\!^{*}$ counts for the two $VCRAE$ tables} \n \\[-12mm]  
   \begin{center}
\begin{tabular}{cccc rrrr l rrrr}
\hline
& &&& \multicolumn{4}{c}{$L=0$, controls} &&\multicolumn{4}{c}{$L=1$, cases} \\
 \cline{5-8} \cline{10-13}
 &&&&  \multicolumn{2}{c}{$E=0$} & \multicolumn{2}{c}{$E=1$} && \multicolumn{2}{c}{$E=0$} & \multicolumn{2}{c}{$E=1$} \\
\multicolumn{4}{c}{$VCRA$ levels}&count& estim.$\!^{*}$&count& estim.$\!^{*}$&&count& estim.$\!^{*}$&count& estim.$\!^{*}$ \\
\hline
     0  &   0  &   0  &   0  &      21&        23.79 &    20  &       19.55  && 2&        0.85 &         4 &         3.29         \\
     1  &   0  &   0  &   0  &      4 &         1.40    &      1 &          1.15&&0 &         0.24  &        1&          0.91          \\
     0  &   1  &   0  &   0  &      8 &          6.85   &       9&          5.63 &&  1   &       0.63&   2&         2.44               \\
     1  &   1  &   0  &   0  &       1&          0.97   &          0  &        0.79   &&1&      1.35&          5&          5.19           \\
     0  &   0  &   1  &   0  &       84&        85.04   &      34 &        35.94   && 4&    2.84&         10&         10.97          \\
     1  &   0  &   1  &   0  &      3&         5.02  &            3&          2.12    &&    2&     2.04 &    6&          7.86          \\ 
     0  &   1  &   1  &   0  &        26&         24.48 &         9&         10.35  &&  2&     2.38 &     12&          9.16         \\ 
     1  &   1  &   1  &   0  &          4&          3.45 &         1&          1.46    && 1  &    1.82 &         6&          7.02        \\
     0  &   0  &   0  &   1  &         4&          5.04  &        28&         29.46   &&  1&   2.85&         11&         11.00         \\
     1  &   0  &   0  &   1  &         1&          0.30   &       1&          1.74   &&   0 &   0.79&          4&          3.06           \\
     0  &   1  &   0  &   1  &          1&        1.45    &      8&          8.48 && 0     &     1.01&     5&          3.91            \\ 
     1  &   1  &   0  &   1  &           0&          0.20&          1&          1.20  &&  2&    2.15  &   9&         8.31            \\
     0  &   0  &   1  &   1  &          21&        18.02&         59&         54.15 && 13&    9.51&    33&        36.68        \\      
     1  &   0  &   1  &   1  &           0 &         1.06  &        3&          3.20    &&  7&    6.82 &  28&         26.29        \\
     0  &   1  &   1  &   1  &          5&         5.19  &       12&         15.59 && 1&        3.80&  15&        14.66          \\ 
     1  &   1  &   1  &   1  &          0&          0.73 &         4&          2.20 &&  5&      2.91  &       11&        11.24           \\ 
  \hline
\end{tabular}
\end{center}\n \\[-6mm]
\small $^{*}$estimates to concentration graphs; for controls in Figure 4b), for cases in Figure 4a)
\vspace{-2mm}
\end{table}    
   \n \\[-4mm]     
   
\noindent{\bf Acknowledgement.}  We thank   Heiko Becher for permitting us to re-analyze the data and
   him, David Cox, Ruth Keogh, and Ivonne Solis-Trapala for their  helpful comments as the work on this paper was progressing. We used Matlab, Stata and R-routines to analyze  the data.

  \renewcommand{\baselinestretch}{0.2}

\end{document}